\documentclass[article]{natureprintstyle}
\usepackage{graphicx}
\usepackage{amsmath}
\usepackage{amssymb} 
\usepackage[hidelinks]{hyperref}
\usepackage{siunitx}
\usepackage{amsfonts}

\providecommand{\blue}[1]{\protect{\color{blue}#1}}

\title{Heterogeneity-stabilized homogeneous states in driven media}
\author{Zachary G. Nicolaou$^1$, Daniel J. Case $^1$, Ernest B. van der Wee$^1$, Michelle M. Driscoll$^1$, and Adilson E. Motter$^{1,2,*}$}

\begin{document}
\maketitle
\begin{affiliations}
\item Department of Physics and Astronomy, Northwestern University, Evanston, Illinois 60208, USA
\item Northwestern Institute on Complex Systems, Northwestern University, Evanston, Illinois 60208, USA\\
*email: \href{mailto:motter@northwestern.edu}{motter@northwestern.edu} \\[1em]\textit{Nature Communications} \textbf{12}, 4486 (2021)\\ DOI: \href{https://doi.org/10.1038/s41467-021-24459-0}{\blue{https://doi.org/10.1038/s41467-021-24459-0}}
\end{affiliations}

\begin{abstract}
\smallskip
Understanding the relationship between symmetry breaking, system properties, and instabilities has been a problem of longstanding scientific interest. Symmetry-breaking instabilities underlie the formation of important patterns in driven systems, but there are many instances in which such instabilities are undesirable.   Using parametric resonance as a model process, here we show that a range of states that would be destabilized by symmetry-breaking instabilities can be preserved and stabilized by the introduction of suitable \textit{system} asymmetry. Because symmetric states are spatially homogeneous and asymmetric systems are spatially heterogeneous, we refer to this effect as heterogeneity-stabilized homogeneity. We illustrate this effect theoretically using driven pendulum array models and demonstrate it experimentally using Faraday wave instabilities. Our results have potential implications for the mitigation of instabilities in engineered systems and the emergence of homogeneous states in natural systems with inherent heterogeneities.
\end{abstract}
\medskip

\begin{figure}[b!]
\includegraphics[width=\columnwidth]{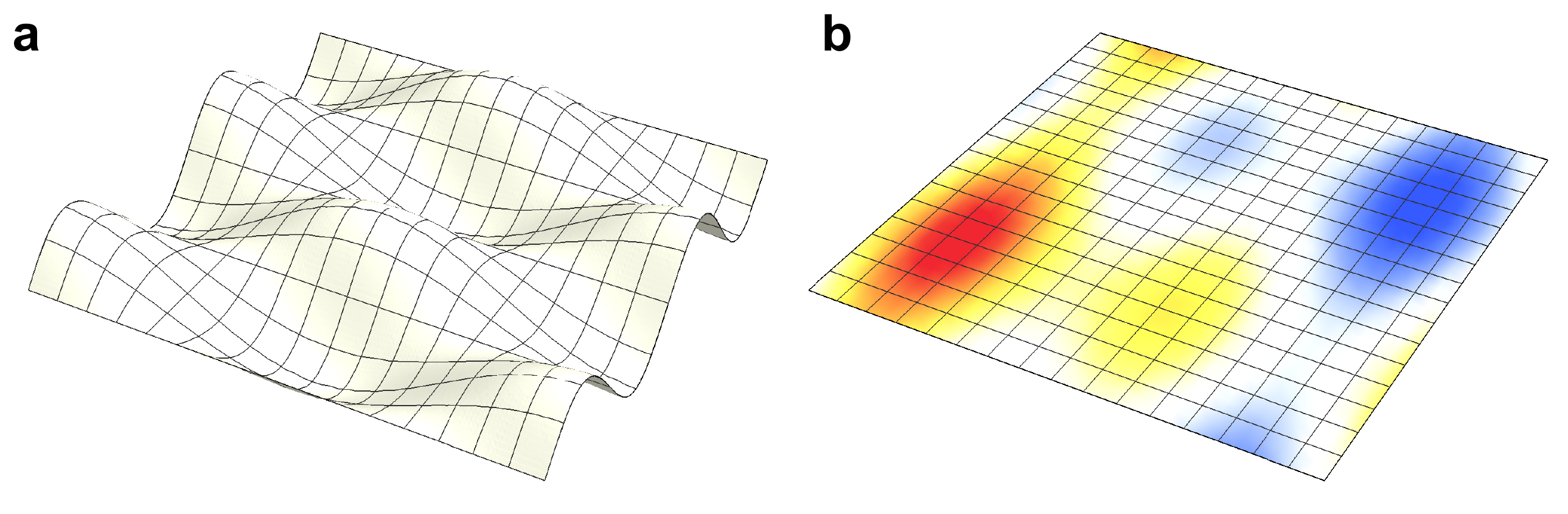}
\caption{\textbf{Conceptual illustration of heterogeneity-stabilized homogeneous states.} \textbf{a} A homogeneous membrane driven by a vertical vibration undergoes a symmetry-breaking instability, resulting in an inhomogeneous (undulated) state. \textbf{b} Introducing heterogeneity in the membrane (represented here by the color) stabilizes the homogeneous (planar) state. The boundary conditions are periodic. \label{fig1}}
\end{figure}

The mitigation of dynamical instabilities is an outstanding problem in diverse areas of science and engineering, ranging from fluid turbulence\cite{2014_swinney}, ecosystem collapses\cite{2019_may}, and cascading  failures\cite{2017_motter} to financial crashes\cite{2007_caldarelli} and resonances in architectural structures\cite{2005_strogatz,2006_comerio}. While progress has been made\cite{1999_sastry,2005_Bacciotti}, the fundamental question of how the properties of a complex system relate to the emergence of instabilities remains largely open. As a result, our ability to mitigate instabilities in such systems remains limited. The problem of preventing instabilities is akin to the problem of stabilizing a desired dynamical state. Of special interest are states that have certain symmetries, such as translational symmetry, time invariance, and permutation symmetry\cite{1967_Prigogine}. For example, an initially planar elastic membrane may become undulated if driven by a uniform transverse periodic force (Fig.~\ref{fig1}a). The question of interest in this case is how to prevent this instability and preserve the fully translationally symmetric state so that the membrane remains flat while driven. For a broad class of systems of interest, symmetric states are guaranteed to exist if the system itself has the same symmetries, but herein lies the rub (and the opportunity): sufficient conditions for the existence of a state are usually neither sufficient nor necessary for the state to be stable.

In this Article, we establish an innovative approach to prevent, delay, or manipulate the onset of symmetry-breaking instabilities in driven complex systems. The approach is based on the realization that states of interest can be preserved {\it and} stabilized  by breaking the symmetry of the system through the introduction of temporally fixed spatial heterogeneities (even when spatial averages are constrained to remain unchanged), as illustrated in Fig.{~\ref{fig1}b}. The underlying phenomenon can be naturally interpreted as the emergence of heterogeneity-stabilized homogeneous states (HSHS). We show that for parametrically-driven systems in particular, the introduction of constrained but appropriately designed heterogeneity provides a general means to stabilize homogeneous states for a wide range of parameter values. Because such homogeneous states emerge without the need for feedback control, HSHS can be exploited to design non-feedback control for complex systems, which has attracted interest in the past for the potential to suppress chaos\cite{1995_Braiman_Lindner}. In the broader context of pattern-forming systems\cite{1952_Turing,1993_Cross_Hohenberg,2001_winfree}, the introduction of disorder other than heterogeneity, such as some forms of noise, has been previously shown to both inhibit instabilities\cite{graham_1982,petrelis_2005} and create multistability\cite{residori_2001,berthet_2003}. Heterogeneity has also been shown to create multistability\cite{coullet_1986,zimmermann_1993} and to shift instability boundaries\cite{pomeau_1993,becker_1994}, but the potential for heterogeneity to stabilize homogeneous states has not been previously recognized.
\begin{figure*}
\begin{center}
\includegraphics[width=2\columnwidth]{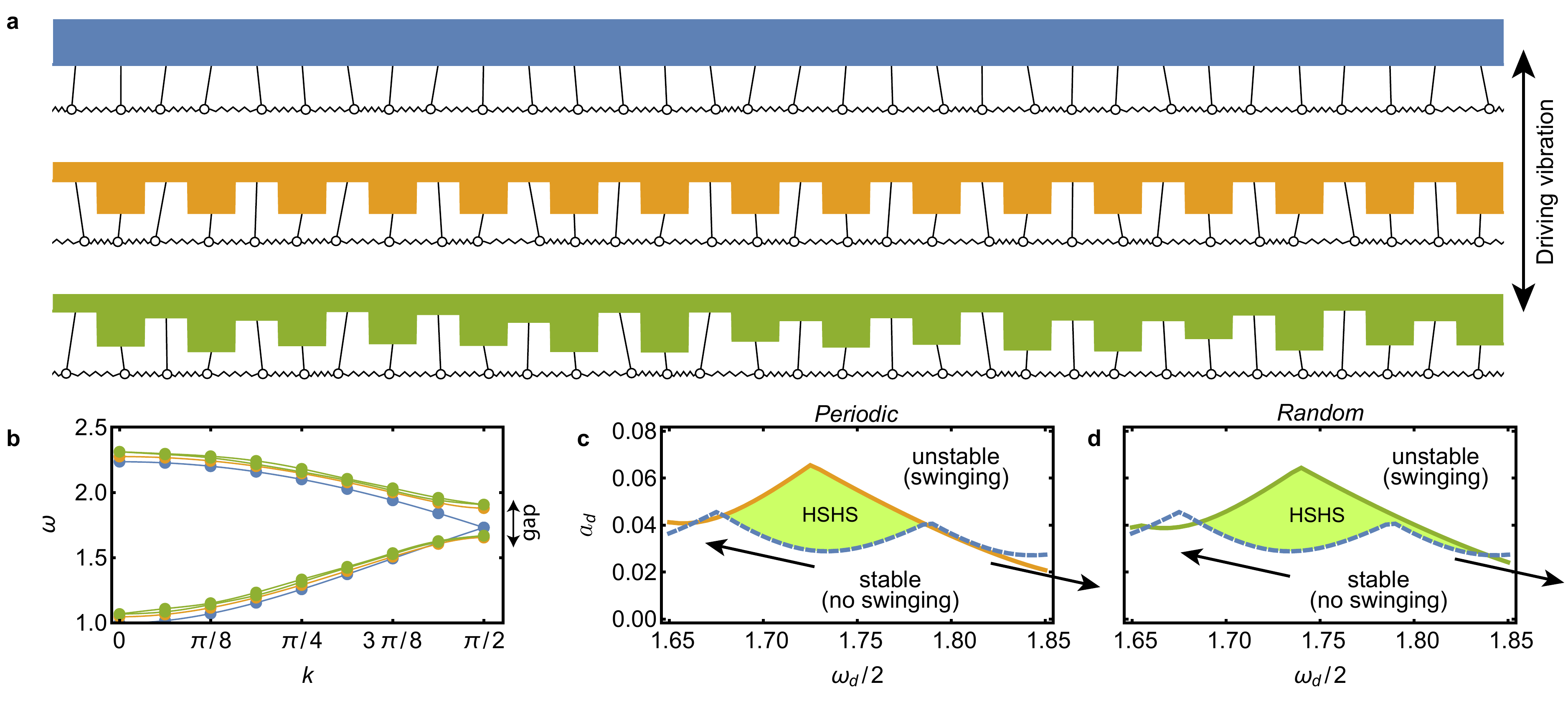}
\caption{\textbf{Heterogeneity-stabilized homogeneous states in arrays of coupled pendula.} \textbf{a} Arrays of coupled pendula with homogeneous support (top), periodic support (middle), and random support (bottom). \textbf{b} Angular frequency $\omega$ vs.\ wavenumber $k$ in the dispersion relation, where the color-corresponding dots show modes for the respective arrays in \textbf{a} and the lines serve as a guide to the eye suggestive of the infinite-pendulum limit. The heterogeneities open a band gap at $k=\pi/2$ around $\omega=1.75$. \textbf{c},\textbf{d} Instability boundaries in the driving amplitude $a_d$ vs.\ driving frequency $\omega_d$ space for the periodic support (\textbf{c}) and the random support (\textbf{d}), with swinging wave motion emerging above the solid lines and with the dashed lines marking this boundary in the homogeneous case. The arrows in \textbf{c},\textbf{d} indicate how instability tongues move in response to the band gap opening, leading to HSHS in the shaded areas. Animations of the pendulum arrays in \textbf{a} are available in Supplementary Movie 1. \label{fig2}}
\end{center}
\end{figure*}

We demonstrate HSHS in two paradigmatic systems. First, we theoretically analyze a driven pendulum array model in order to illustrate the following general mechanism for the generation of HSHS. The introduction of heterogeneities can create a band gap in the dispersion relation, which follows from the different ways in which the short- and long-wavelength modes are impacted by heterogeneity. This band gap creates a region of the driving amplitude vs.\ driving frequency parameter space with no modes that can be resonantly excited by the driving, leading to a stabilization of the homogeneous state in this region. We then turn to the demonstration of HSHS in Faraday wave instabilities, which are important in their own right and are representative of a large class of instabilities in driven systems\cite{1954_Benjamin_Ursell, 2017_Gallaire_Brun}. Faraday waves are standing waves that emerge on the surface of a liquid in a vertically vibrating container. They appear above threshold values of driving frequency and amplitude that mark the instability boundary where the flat fluid surface becomes unstable. Faraday wave experiments are usually performed using a flat substrate for the bottom of the container, which guarantees that the system is spatially homogeneous and thus has translational symmetry (up to boundary effects). This symmetry is spontaneously broken by the standing waves that are created by parametric resonance above the instability boundary. We consider how Faraday instabilities are impacted by symmetry-broken geometries, defined by heterogeneous (i.e., non-flat) substrates. Past studies have considered the possibility of localization resulting from small corrugations in the container\cite{2001_Osipov_Garcia, 2015_Weidman_Howard, 2016_Feng_Stone} or through localized driving forces\cite{2019_Urra_GarciaNustes}, but we stress that the potential for HSHS has not been considered in this literature. Here, we show how heterogeneity from more general substrate geometries can impact the onset of Faraday wave instabilities, which allows us to experimentally demonstrate the existence of HSHS for both sinusoidal and random substrates with suitably large heterogeneity.

\section*{Results\\}
 \subsection{Mitigating instabilities in a model system~}
 It is instructive to first consider an instability in a discrete model system consisting of an array of identically-coupled identical pendula, as shown in Fig.~\ref{fig2}a (blue). Each pendulum in this model experiences gravitational forces and is coupled to its nearest neighbors via linear springs. We employ periodic boundary conditions, so that the system is completely symmetric with respect to translations by one lattice site. Similar models of coupled oscillators have been of recent interest in the study of classical time crystals\cite{2019_Heugel,2020_Yao_Zaletel,2021_Nicolaou}. When the system is driven by a vertical vibration, defined by a driving frequency $\omega_d$ and a driving amplitude $a_d$, symmetry is spontaneously broken and the pendula start to swing sideways as the driving frequency and/or amplitude is increased above an instability boundary.

To understand how to mitigate this instability, we first focus on the homogeneous system in the absence of driving. In this case, the dispersion relation, which describes the response frequency $\omega$ as a function of the wavenumber $k$, can be used to decompose general disturbances into wave modes (see`` Methods''). The dispersion relation for the homogeneous pendulum array (blue line in Fig.~\ref{fig2}b) consists of two branches corresponding to, respectively, short- and long-wavelength modes, which merge at $k=\pi/2$ for an array of $N$ pendula with pivots equally spaced horizontally (the number of pendula is assumed to be even to facilitate our subsequent analysis for unit cells of size two). When the system is driven at a frequency $\omega_d$, on the other hand, the instability boundary of the (steady) homogeneous state can be determined through a generalized eigenvalue problem for the critical driving amplitude $a_d^{~*}$,
\begin{equation}
\label{pendulumeigen}
\sum_\nu \tilde{A}_{ \mu \nu} \tilde{\theta}_\nu =a_d^{~*} \sum_\nu \tilde{B}_{\mu \nu} \tilde{\theta}_\nu,
\end{equation}
where $\tilde{\theta}_\nu$ are Floquet mode amplitudes and $\tilde{A}_{\mu \nu}$ and $\tilde{B}_{\mu \nu}$ are matrices encoding the linearized dynamics (see ``Methods''). For driving amplitudes above the critical value, the swinging motion of the pendula that results from spontaneous symmetry breaking is decomposed (according to the driving frequency) into excitations of individual modes in the dispersion relation of the undriven system. These instabilities are subharmonic{\cite{2004_Braun}} and correspond to the wave modes oscillating at one half the driving frequency, which are marked as dots for $N=32$ in Fig.~\ref{fig2}b.

We propose that such instabilities can be mitigated by introducing spatial heterogeneity. This follows from two observations. First, when the dispersion relation has no resonant wave modes (corresponding to half the driving frequency), larger driving amplitudes are required to create instabilities induced by exciting non-resonant modes. Thus, we expect that a general approach to mitigate instabilities will be by shifting the instability boundary through the creation of band gaps in the dispersion relation, where modes are not easily excited. Second, we note that given a particular length scale, the dispersion relation can be naturally divided into wave branches whose dynamics differ from each other on that scale. In the homogeneous system, these wave branches meet at points of degeneracy. When heterogeneity of that length scale is introduced, the branches will tend to respond differently and the degeneracies will be lifted, effectively creating band gaps by separating the branches. As we show explicitly for one-dimensional lattices in ``Methods'', this band gap opening mechanism is generic for periodic heterogeneity and provides a systematic means to introduce desirable band gaps through tunable heterogeneities. We suggest that the method that results from combining these two observations is applicable to parametric instabilities in general.

We can now demonstrate this approach in the pendulum array model by introducing temporally fixed heterogeneity in the lengths of the pendula. We implement heterogeneity by spatially varying the height of the support ceiling while keeping the rest position of the pendulum bobs leveled. To avoid conflating effects, everything else is kept unchanged (including the average length of the pendula). We consider both periodic and random configurations of pendulum lengths, as shown in Fig.~\ref{fig2}a (orange and green, respectively). The system's translational symmetry is partially broken for the periodic configuration and fully broken for the random one. As shown in Fig.~\ref{fig2}b (color coded as in Fig.~\ref{fig2}a), these heterogeneities open band gaps in the dispersion relation, which is gapless in the homogeneous case. While we have introduced heterogeneity in the pendulum lengths to motivate our subsequent Faraday wave experiments, general considerations of wave-lattice interactions imply that a band gap will appear generically in periodic configurations regardless of how the heterogeneity is implemented (see ``Methods''). As an application of this general result, we show that heterogeneity in the masses of the bobs is also capable of stabilizing homogeneous states in a driven pendulum array (see Supplementary Discussion 1 and Supplementary Fig. 1).

The orange and green lines in Fig.~\ref{fig2}c,d show the instability boundaries predicted from the Floquet analysis defined by \eqref{pendulumeigen}, with the dashed blue lines showing the corresponding boundary for the flat configuration. The shaded areas between the curves mark the HSHS regions, where the system is stabilized by the heterogeneity. Unlike previous studies focused on exploring the balance between two controllable asymmetries to restore features of symmetry breaking\cite{2020_Garbin} or on suppressing chaos in favor of symmetry-broken periodic states\cite{1999_Shew}, here the band gap created by appropriate heterogeneity suppresses the parametric instability, resulting in a homogeneous steady state.  The instability suppression occurs for driving frequencies around $\omega_d/2= 1.75$ since the instabilities are subharmonic and the band gap opening is around $\omega = 1.75$. This opening is determined by the heterogeneity magnitude and length scale, and it can be manipulated by varying these parameters. The areas below the instability boundaries for the periodic and random case in Fig.~\ref{fig2}c,d are thus larger than for the homogeneous case (by $28.7\%$ and $30.4\%$, respectively). In both cases, the results confirm our hypothesis that the band gap opening leads to stabilization of homogeneous states by moving apart subharmonic instability tongues near the band gap. We also note that the random case has an even larger gap (and hence, a larger region of HSHS) than the periodic case for the same magnitude of heterogeneity. An analogous band-gap widening effect has been recently reported in the design of metamaterials with spatial disorder\cite{2019_Celli}. Altogether, these results illustrate that band gaps emerging from both periodic and non-periodic heterogeneities can influence the instability boundaries in the driven system, leading to HSHS in a large portion of parameter space.

In addition to stabilizing homogeneous states, the introduction of heterogeneity can have other consequences for the dynamics of the driven pendulum array. We find that while a homogeneous array exhibits either a steady homogeneous or a swinging stable state (depending on the driving frequency and amplitude), heterogeneous arrays can also exhibit multistability for some driving parameters. Within parts of the region of HSHS, it is possible for finite-size perturbations to excite the steady homogeneous state into states with localized swinging motion, in which the swinging motion is mostly confined to a few neighboring pendula (see Supplementary Movie 1). Such states have been shown to appear in connection with band gaps in other systems and are known as gap solitons\cite{2018_PonedeL_Knobloch}. These gap solitons coexist with the homogeneous state in the heterogeneous pendulum array, but we find that large perturbations are required to excite them (see Supplementary Discussion 2 and Supplementary Fig. 2). 

\subsection{Demonstrating HSHS in a fluid experiment~}
\begin{figure*}
\begin{center}
\includegraphics[width=2\columnwidth]{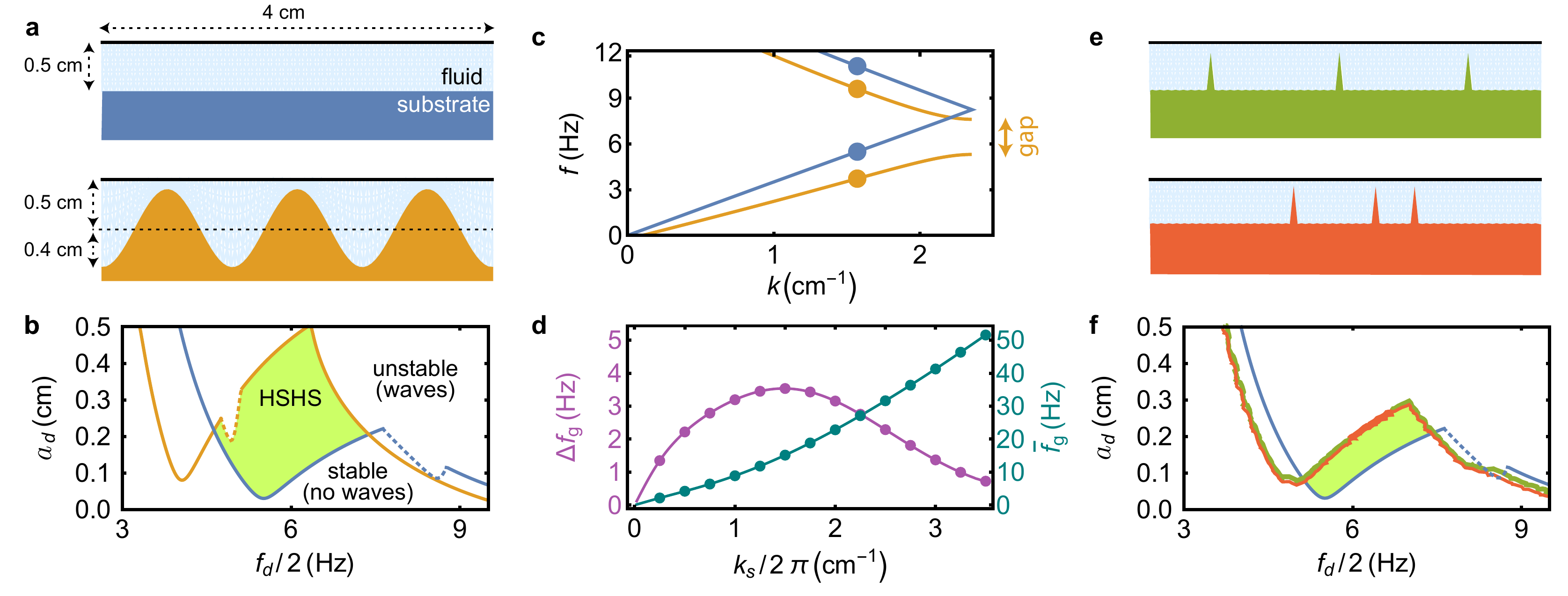}
\caption{\textbf{Faraday instabilities in one dimension with periodic boundary conditions.} \textbf{a} Flat (blue) and sinusoidal (orange) substrate geometries. \textbf{b} Instability boundaries in the driving amplitude vs.\ driving frequency space for the color-corresponding substrates in \textbf{a} showing the subharmonic (continuous lines) and harmonic ({dotted} lines) instability boundaries as well as the resulting HSHS region (shaded area). \textbf{c} Frequency vs.\ wavenumber in the dispersion relation for the substrates in \textbf{a} (dots) and for the respective infinite-length systems (lines). \textbf{d} Frequency gap $\Delta f_g$ and mean frequency $\bar{f_g}$ of the modes at the band gap as a function of the substrate wavenumber $k_s$, showing wavenumbers that meet periodic boundary requirements for the substrates shown in \textbf{a} (dots). \textbf{e} Substrate geometries with evenly (green) and randomly (red) spaced sharp peaks. \textbf{f} Instability boundaries as in \textbf{b} for the flat substrate in \textbf{a} and the sharp-peak substrates in \textbf{e}.
\label{fig3}}
\end{center}
\end{figure*}
We now turn to the demonstration of HSHS in Faraday instabilities, which we recall are characterized by the emergence of standing waves on the surface of fluid-filled containers driven by vertical vibrations. Similarly to the case of the pendulum array model, these instabilities occur when the system is driven above an instability boundary in the driving amplitude $a_d$ vs.\ driving frequency $f_d$ parameter space (here we use frequencies $f$ rather than angular frequencies $\omega$ to facilitate comparison with the experiments). To implement the heterogeneity, we consider containers with rectangular cross sections and with spatially varying depths. In all containers, we maintain an identical cross section and an identical total volume of fluid.

In order to design our experiments, we first model and simulate HSHS for the Faraday instability. For simplicity, we assume the fluid is incompressible and inviscid, which adequately describes experiments with water, although our analysis can be extended to include viscid effects\cite{1994_Kumar_Tuckerman}. Our experiments below are conducted using water in containers measuring $4$ \si{cm} in length by $1$ \si{cm} in width and with an average depth of $0.5$ \si{cm}. The onset of instability in these large-aspect-ratio containers can be approximated using a one-dimensional model (of the same length and depth) with periodic boundary conditions. The use of periodic boundary conditions is desirable in order to show that the effect does not result from asymmetries caused by the end walls (and below we also show that the effect persists for other boundary conditions). For periodic boundary conditions, the eigenvalue problem analogous to \eqref{pendulumeigen} for Faraday instabilities derived from the Navier-Stokes equations is
\begin{equation}
\label{faradayeigen}
\sum_\nu \tilde{C}_{\mu \nu} \tilde{h}_\nu = a_d^{~*} \sum_\nu \tilde{D}_{\mu \nu} \tilde{h}_\nu,
\end{equation}
where $\tilde{h}_\nu$ is a Floquet-Fourier mode amplitude and $\tilde{C}_{\mu \nu}$ and $\tilde{D}_{\mu \nu}$ are linearization matrices. The additional Fourier decomposition in this analysis, not present in the pendulum array case, arises because the spatial variables must be converted to discrete modes to carry out the Floquet analysis (see ``Methods'').

Figure \ref{fig3} shows the results of simulations of the fluid system for various substrate shapes. We compare the flat substrate to a sinusoidal substrate with wavenumber $k_s$ and amplitude $a_s$, as shown in Fig.~\ref{fig3}a for $k_s=3\pi/2$ \si{cm^{-1}} and $a_s=0.4$ \si{cm}. The blue and orange lines in Fig.~\ref{fig3}b show the instability boundary for these substrates, as determined by \eqref{faradayeigen}. While most of the instabilities are subharmonic, small harmonic instability tongues (corresponding to oscillations at the same frequency as the driving) protrude for specific driving frequencies, which are shown by the dotted lines. The instability boundaries (both harmonic and subharmonic) determined by \eqref{faradayeigen} agree exactly with those determined through direct finite element simulations (see ``Methods''), which confirms the consistency of our analysis. The shaded area in Fig.~\ref{fig3}b corresponds to HSHS, indicating that substrate heterogeneity can indeed stabilize Faraday instabilities in a large portion of the parameter space. The heterogeneous geometry in Fig.~\ref{fig3}a has a fluid layer of $1$ \si{mm} between the substrate peaks and the undisturbed fluid surface, which is adequate for the wavelength modes considered to be significantly affected by substrate heterogeneity given that the dispersion relation depends exponentially on the minimum fluid thickness (see ``Methods''). We have checked, however, that the observed HSHS are not a result of simple compartmentalization of the fluid between successive peaks of the substrate by considering substrates defined by an equal number of sharp peaks with the same minimal fluid thickness for the same fluid volume (Fig.~\ref{fig3}e-f). Importantly, sharp peaks do a substantially poorer job of stabilizing the flat surface than a sinusoidal substrate, and the result is largely independent of the positions of the peaks.

While we focus on specific substrates in our experiments, we note that the same results follow for containers with different lengths and different $k_s$. Figure \ref{fig3}c shows the dispersion relation for systems of infinite length with the same $k_s$ as in Fig.~\ref{fig3}a. For the finite-length system we consider, the first two excited modes (dots in Fig.~\ref{fig3}c) lie on different branches and are further separated as the gap opens, leading to the observed HSHS. This effect persists for containers of arbitrarily large length, since there are no modes available which can be easily excited within the band gap. The harmonic instabilities in Fig.~\ref{fig3}b (dotted lines), on the other hand, do not occur at the instability onset for flat substrates in sufficiently long containers. Instead, subharmonic instabilities of longer wavelength (which would not fit within the $4$ \si{cm} bounded container) occur at lower driving amplitudes unless the system is driven by non-sinusoidal vibrations\cite{2006_Huepe_Silber}. For the sinusoidal substrate, however, the harmonic instabilities can persist for sinusoidal vibrations even in the infinite length limit since the subharmonic modes become difficult to excite near the band-gap frequency. As the substrate wavenumber varies, the position and size of the band gap vary as well. Figure \ref{fig3}d shows the frequency gap $\Delta f_g$ and the mean frequency $\bar{f_g}$ for the band gap at $k=k_s/2$ as a function of $k_s$. Shorter substrate wavelengths correspond to a band gap (and thus to HSHS) at higher frequencies, with the largest band gap appearing for $k_s/2\pi \approx 1.4$.

Having established these theoretical predictions, we now assess them experimentally. We consider flat and sinusoidal substrates as well as randomly generated substrate shapes representing more general heterogeneities. Because larger frequencies are more easily accessible in experiments, we used a larger substrate wavenumber of $k_s/2\pi=1$ \si{cm^{-1}} for the sinusoidal substrate (see ``Methods''). For the random substrates, the parameter space of substrate heterogeneities is very large---infinite, in fact. To generate random substrates, we sample this space by taking the Fourier coefficients of the first $12$ modes as random Gaussian variables (with the mean inversely proportional to the Fourier index). We then employed the width of the range of stabilized frequencies between $10$ \si{Hz} and $14$ \si{Hz} for a fixed driving acceleration of $a_d \times (2\pi f_d)^2=0.8 g$ as a computationally affordable proxy for the stabilized area in the driving amplitude vs.\ driving frequency space ($g=9.8$ \si{m/s^2} is the gravitational acceleration). We optimized this value over $500$ randomly generated substrates to select the random substrate used in experiments. The flat substrate, the sinusoidal substrate, and the selected random substrate were then 3D printed to use in our experiments (see ``Methods'').

Figure \ref{fig4}a shows the instability boundaries for these substrates, as determined by our experiments (and in agreement with our numerical predictions). The experiments reveal a large region of HSHS in which the flat surface is stable for the sinusoidal or random substrates but not for the flat substrate, as indicated by the shaded area in the driving amplitude vs.\ driving frequency space. Snapshots of the HSHS are shown in Fig.~\ref{fig4}b,c. While the selected random substrate stabilizes a slightly smaller region than the sinusoidal one, we do not expect the sinusoidal case to be generally optimal in terms of the area of the stabilized region{,} given that this is not the case for the pendulum system. Still, the selected random heterogeneity stabilizes a large portion of parameter space and demonstrates that this effect is not specific to sinusoidal or approximately periodic substrates. These experiments clearly show that the band gap opening mechanism behind HSHS is general and robust enough to be observed in realistic systems.

\section*{Discussion\\}
Our demonstration of scenarios in which parametric instabilities can be mitigated by introducing heterogeneity reveals a new phenomenon in complex spatiotemporal systems. The mechanism giving rise to this phenomenon stems from the formation of a band gap in the dispersion relation, which produces a shifting of the instability tongues in parameter space. Our Faraday wave experiments explicitly show that the band gap opening that leads to the desirable stabilization of homogeneous states can indeed be implemented with the introduction of appropriately designed system heterogeneity. These results show the existence of homogeneous states that not only persist in spite of heterogeneity but are also promoted by it.

 \begin{figure*}
 \begin{center}
\includegraphics[width=2\columnwidth]{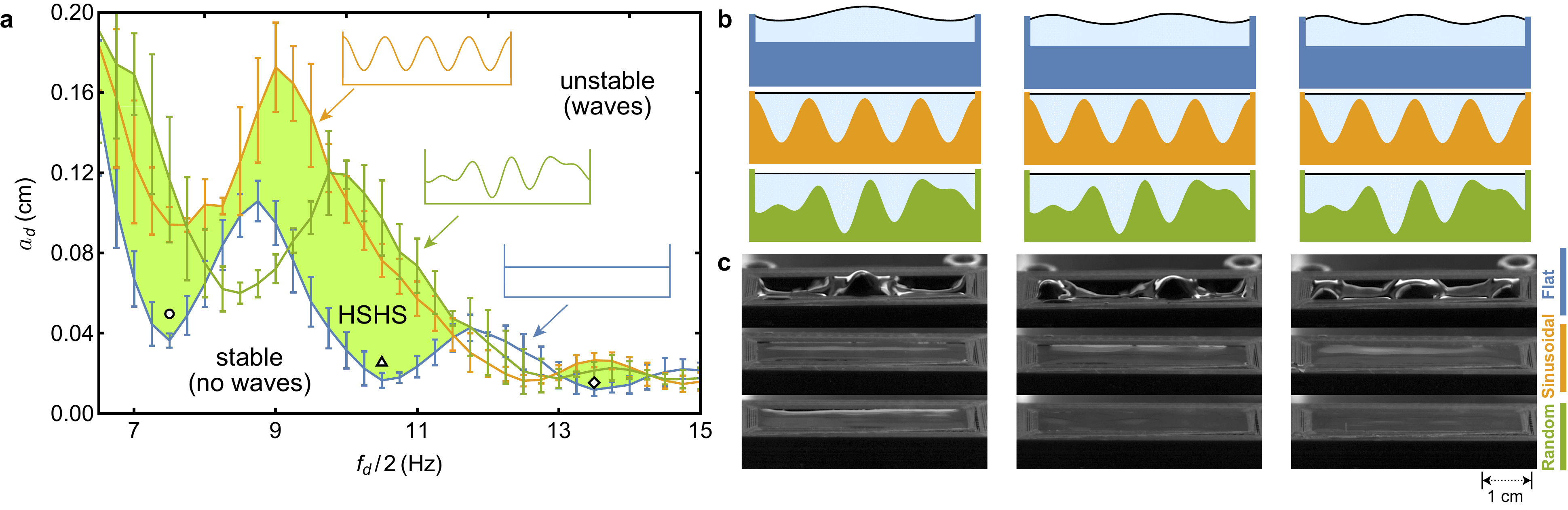}
\caption{\textbf{Heterogeneity-stabilized homogeneous states in Faraday instability experiments.} \textbf{a} Experimental instability boundaries in the driving amplitude vs.\ driving frequency space for the flat substrate (blue), the sinusoidal substrate (orange), and the selected random substrate (green), as indicated by the corresponding colors insets. The error bars show the standard error from 4 measurements of the critical driving amplitude at each frequency, and the shaded area indicates parameters for which the shifting of the instability tongues leads to HSHS. \textbf{b},\textbf{c} Numerical (\textbf{b}) and experimental (\textbf{c}) snapshots for HSHS for the three substrates (corresponding substrates are marked by the same color). The columns in \textbf{b}, \textbf{c} correspond to the driving parameters marked by the circle (left), the triangle (middle), and the diamond (right) in \textbf{a}. Videos of the experiments in \textbf{c} are available in Supplementary Movie 2. \label{fig4}}
\end{center}
\end{figure*}

The emergence of HSHS is also intimately related to the concept of symmetry. It shows that unstable symmetric states may continue to exist and, most importantly, become stable when the corresponding symmetry of the system itself is explicitly broken. This phenomenon can thus be interpreted as an analog in continuous media of scenarios in which stable synchronization requires system asymmetry\cite{2016_nishikawa} recently observed in discrete systems, and which have been proposed to emerge from network interactions\cite{2019_hart,2020_molnar}. Our results establish that this previously unrecognized symmetry phenomenon occurs in continuous media and can be used to suppress parametric instabilities. This work also contributes a new insight into the broader literature on the relation between disorder and pattern formation, which has included studies ranging from Anderson localization\cite{Lagendijk_Wiersma_2009} to the impact of disorder\cite{1995_Braiman_Lindner} and impurities\cite{1998_Gavrielides_Tsironis,2000_Alexeeva_Barashenkov} on spatiotemporal chaos and soliton dynamics. Ultimately, our results demonstrate a new approach to suppress instabilities in parametrically driven media by manipulating system parameters. Such an approach is materially different from prior efforts to prevent instabilities based on the explicit control of the system variables, which is often limited by one's ability to actuate the required variables in real time.

We suggest that the HSHS effect has the potential to be a general phenomenon occurring in other systems undergoing instabilities due to mode excitation. In those systems too, we expect to have the opportunity to suppress instabilities through the manipulation of band gaps created by the introduction of heterogeneity. For example, complex instabilities are known to emerge in driven elastic membranes\cite{2004_Awrejcewicz}, but elastic composites with periodic heterogeneity have been shown to exhibit band gaps\cite{2011_Andrianov}, which our results suggest could be designed to mitigate these instabilities. Importantly, this approach can benefit from previous studies on creating and manipulating band gaps for different purposes, including the literature on topological edge states in both discrete systems\cite{2015_Pai_Bertoldi,2018_Mitchell_Irvine,2019_Palmero} and continuous media\cite{2015_Yang_Zhang,2018_PonedeL_Knobloch}. The broader implications of this work for future studies include the modeling of heterogeneity in natural systems, such as in describing its coevolution with homogeneous states, and the mitigation of instabilities in experimental and man-made systems.

\begin{methods}
\section*{Pendulum array theory and numerics.}
\ The state of each pendulum in the array is described by the angle $\theta_i$ with the vertical direction, where $1\leq i \leq N$ is the pendulum index. In the accelerated reference frame of the moving ceiling, the equations of motion are
\begin{align}
\label{springs}
M L_i \ddot{\theta}_i &= -\eta L_i \dot{\theta}_i-M \big[ g-a_d \omega_d^2 \cos(\omega_d t) \big]\sin(\theta_i) \nonumber \\
& \quad +\kappa L_{i+1} \sin(\theta_{i+1}-\theta_i) +\kappa L_{i-1} \sin(\theta_{i-1}-\theta_i) \nonumber \\
&\quad + \kappa\left[L_{i+1}+L_{i-1}-2L_i\right] \sin(\theta_i),
\end{align}
where $t$ is time, overdots denote time derivatives, $M$ is the pendulum mass, $L_i$ is the length of the $i$th pendulum, $\kappa$ is the spring constant (springs are assumed to have zero unstretched length), $g$ is the gravitational constant, $\eta$ is the damping coefficient, $\omega_d=2\pi f_d$ is the angular frequency of the vertical vibration, and $a_d$ is the amplitude of the vibration. The pendulum lengths are defined as $L_i=\overline{L}$ for the homogeneous support and $L_i = \overline{L} + (-1)^i \Delta$ for the periodic configuration. We call the average square deviation of the pendulum lengths, $\langle \left(L_i-\overline{L}\right)^2 \rangle$, the heterogeneity magnitude, which in this case reduces to $ \left(L_i-\overline{L}\right)^2=\Delta^2$. For the random case, $L_i =\overline{L} + \delta_i$ for $\delta_i$ sampled from a uniform distribution with mean $(-1)^i\Delta'$ and variance $\gamma^2 \left(\Delta'\right)^2/3$, where $\Delta' \equiv \Delta/\sqrt{1+\gamma^2/3}$. This distribution guarantees that the average pendulum length remains unaltered (i.e., $\langle L_i \rangle = \overline{L}$) and that the heterogeneity magnitude is the same as in the periodic support (i.e., $\langle \left(L_i-\overline{L}\right)^2 \rangle = \Delta^2$). In our simulations, we take $N=32$, $\overline{L}=1$, $\gamma=0.5$, and $\Delta=0.35$ and assume that all quantities are nondimensionalized.

We linearize the equations of motion using the Floquet ansatz
\begin{equation}
\theta_i ={\mathrm e}^{st} \sum_m \hat{\theta}_{i m} \mathrm{e}^{\mathrm{j} m \omega_d t},
\label{fourier}
\end{equation}
where $\hat{\theta}_{i m}$ are Fourier mode amplitudes, $s=\beta+\mathrm{j}\omega_d \epsilon$ is the Floquet exponent with growth rate $\beta$ and response frequency ratio $\epsilon=\omega/\omega_d$, and $\mathrm{j}$ is the imaginary unit. Inserting \eqref{fourier} into \eqref{springs} results in
\begin{equation}
\label{eigen}
\sum_i \sum_m {A}^{i m}_{j n}\hat{\theta}_{i m} = a_d\sum_i \sum_m {B}^{i m}_{j n} \hat{\theta}_{i m}, \\
\end{equation}
where
\begin{align}
{A}^{i m}_{j n} &= L_j\left[ -M\omega_d^2(s+n)^2+\mathrm{j} \omega_d\eta(s+n)+2\kappa \right]\delta^i_j\delta^m_n \nonumber \\
&\quad -\left[\kappa L_{j+1} \delta^i_{j+1} + \kappa L_{j-1} \delta^i_{j-1} - M g \delta^i_j \right]\delta^m_n, \\
{B}^{i m}_{j n} &= \frac{M}{2}\omega_d^2\left[\delta^m_{n+1} + \delta^m_{n-1} \right]\delta^i_j,
\end{align}
and $\delta^i_j$ is the Kronecker delta. The integers $i$ and $j$ run over the pendulum indices with periodic boundary conditions, while $n$ and $m$ run over the Floquet modes. We map indices $\mu \in \mathbb{Z}$ and $\nu \in \mathbb{Z}$ to pairs of indices $(i(\mu),m(\mu))\in \mathbb{Z}^2$ and $(j(\nu),n(\nu))\in \mathbb{Z}^2$ using any convenient bijection from $\mathbb{Z}$ to $\mathbb{Z}^2$ to define $\tilde{A}_{\mu \nu} \equiv A^{i(\mu)\ m(\mu)}_{j(\nu)\ n(\nu)}$, $\tilde{B}_{\mu \nu} \equiv B^{i(\mu)\ m(\mu)}_{j(\nu)\ n(\nu)}$, and $\tilde{\theta}_{\mu} \equiv \hat{\theta}_{i(\mu)\ m(\mu)}$. Because the problem in the undriven case becomes diagonal in the Floquet space, we can set $n=m=0$ and consider the spectrum of $A_{j0}^{i0}$ to determine the dispersion relation numerically for any configuration of pendulum lengths. In the periodic case, the translational symmetry implies that the eigenfunctions will be sinusoidal in space with wavenumber $k$, resulting in an analytic expression for the dispersion relation given by
\begin{equation}
\left|
\begin{array}{cc}
-M\omega ^2+\mathrm{j} \eta \omega +2 \kappa +\frac{Mg}{\Delta +1} & \frac{2 \kappa (1-\Delta ) \cos k}{\Delta +1} \\
\frac{2 \kappa (\Delta +1) \cos k}{1-\Delta} & -M\omega ^2+\mathrm{j} \eta \omega +2 \kappa +\frac{Mg}{1-\Delta} \\
\end{array}
\right| = 0,
\end{equation}
where $| \cdot |$ is the determinant.

To derive \eqref{pendulumeigen} for the instability boundary in the presence of driving, we set the growth rate to $\beta=0$ and the acceleration to the critical value $a_d^{~*}$ in \eqref{eigen} for a specified frequency ratio $\epsilon$. The critical driving amplitude is then given by the smallest solution to the generalized eigenvalue problem for each driving frequency, which is found numerically by truncating after the first $7$ Floquet modes, so that $-3 \leq n,m \leq 3$. The frequency ratio can be constrained to the first Brillouin zone $-1/2\leq \epsilon\leq 1/2$, since changes in the response frequency $\omega = \epsilon \omega_d$ by a multiple of $\omega_d$ can always be absorbed into the coefficients $\hat{\theta}_{i m}$. We can thus set $\epsilon=0$ and $\epsilon=\pm 1/2$ to find all harmonic and subharmonic instabilities, respectively, as there are no real eigenvalue solutions $a_d^{~*}$ for other $\epsilon$ (but note that anharmonic instabilities with irrational $\epsilon$ can also occur in other cases\cite{2021_Nicolaou}). We note that for the homogeneous system (and non-generically for heterogeneous systems), translationally invariant instabilities in the form of homogeneous swinging states can occur by exciting the steady homogeneous state for $k=0$. Such states emerge from symmetry breaking instabilities of the time translational invariance instead and, thus, when accounting for all symmetries, they are not considered to constitute HSHS. Owing to fluid incompressibility, no time-dependent homogeneous states can occur in our Faraday instability experiments.

 The Floquet analysis was verified to agree with direct numerical integration of \eqref{springs} performed using the Mathematica implementation of the LSODA adaptive time-stepping, stiffness-switching integrator. In these simulations, the Euclidean norm of the vector of angles $\theta_i$ is fit in time to determine the growth rate $\beta$ and the instability frequency $\omega$ is extracted by examining the peaks in the Fourier transform of the de-trended norm.

 \section*{Band gaps in more general models.}
\ Here we generalize the mechanism for HSHS in parametrically-driven systems by describing how periodic heterogeneities of arbitrary form lead to the formation of band gaps in general extended systems with periodic boundary conditions. Crucially, by employing a parameterization of wave modes that correctly reflects the reduced symmetry of the heterogeneous system, the dispersion relation divides itself naturally into branches that meet at degenerate points. These branches are distinguished from each other by the dynamics of the modes in the homogeneous system over the length scale defined by the heterogeneity. In the presence of the heterogeneity, the branches respond differently, and the degeneracies are generically lifted. The latter underlies the formation of band gaps in the dispersion relation.

For concreteness, we focus on the onset of small-amplitude waves in one-dimensional lattice models with arbitrary short- and long-range coupling, but the following arguments imply the opening of band gaps at critical points of the first Brillouin zone more generally. We consider linear second-order models given by
\begin{equation}
{M \ddot{\psi}_n+ \eta \dot{\psi}_n + \sum_{m=1}^N G_{nm} \psi_m + \lambda \Delta_{n} \Big( \{ \psi_j, \dot{\psi}_j, \ddot{\psi}_j \}_{j=1}^{N} \Big) =0,
\label{homogeneous}
}\end{equation}
where $\psi_n$ for $n=1,2,\cdots,N$ is a state variable at each lattice site, $M$ is a mass parameter, $\eta$ is a damping parameter, $G_{nm}$ denotes the elements of a coupling matrix, and $\lambda\Delta_n$ describes a spatial heterogeneity with overall amplitude $\lambda$. We take the translational-invariance assumption that the components of the coupling matrix depend only on the distance $\ell=\mathrm{min} ( n\!-\!m{~\mathrm{mod}~}N,\, m\!-\!n{~\mathrm{mod}~}N )$ between the lattice points, i.e., $G_{nm}={G}(\ell)$ for some function $G$, which implies that the system is homogeneous for $\lambda=0$. The linear form of \eqref{homogeneous} for $\lambda=0$ suggests decomposition of the solution into wave modes. In such a homogeneous case, the translational-invariance assumption guarantees that the wave modes are given by $\psi_n=\mathrm{e}^{\beta_0 t+\mathrm{j}\omega_0 t} \mathrm{e}^{\mathrm{j}qn}$, parameterized by a wavenumber $q\in[-\pi,\pi]$. The frequency $\omega_0(q)$ and growth rate $\beta_0(q)$ of each mode in the homogeneous system are given by the dispersion relation $M\big(\beta_0(q)+\mathrm{j}\omega_0(q)\big)^2+\eta\big(\beta_0(q)+\mathrm{j}\omega_0(q)\big) +\widehat{G}(q)=0$, where ${\widehat{G}(q)=G(0)+2\sum_{\ell=1}^{\lfloor N/2 \rfloor} \cos(q\ell){G}(\ell)}$ is the Fourier transform of the coupling function.

We next consider the impact of heterogeneities of the form
\begin{equation}
\Delta_{n} = \sum_m\left( M'_{nm}\ddot{\psi}_m + \eta'_{nm}\dot{\psi}_m + G'_{nm} \psi_m\right),
\label{heterogeneous}
\end{equation}
and we take the heterogeneity as a small perturbation in \eqref{homogeneous} (i.e., $\lambda \ll 1$). For simplicity, we assume that the heterogeneity possesses a coarser translational invariance than the original problem. In particular, let $T^m_{n} = \delta^m_{n+N_h}$ denote the components of the linear operator that translates the system by $N_h$ lattice points, where $\delta^{m}_n$ is the Kronecker delta. We assume that $M'_{nm}$, $\eta'_{nm}$, and $G'_{nm}$ all commute with $T^m_{n}$, so that there are $N_h$ sites in the unit cell of the heterogeneous system. The translational symmetry of \eqref{heterogeneous} implies that we can find wave modes for \eqref{homogeneous} that simultaneously diagonalize $T^m_{n}$, which has eigenvalues $\mathrm{e}^{\mathrm{j}qN_h}$ and corresponding eigenvectors $\mathbf{v}^q$ with components $v^q_n=\mathrm{e}^{\mathrm{j}qn}$.

We now determine perturbative frequencies and growth rates of the wave modes in \eqref{homogeneous}. Note first that the translational invariance of the homogeneous system implies that $\omega_0(q)$ and $\beta_0(q)$ are even functions. Thus, the modes given by $q$ and $-q$ are degenerate in frequency and growth rate. However, the corresponding eigenvalues of $T^m_{n}$, namely $\mathrm{e}^{\pm\mathrm{j}qN_h}$, are distinct unless $qN_h \mathrm{~mod~} {\pi} = 0$. It follows that for $qN_h \mathrm{~mod~} \pi \neq 0$, the $q$ and $-q$ modes cannot mix and nondegenerate perturbation theory can be applied. Thus, the frequency $\omega$ and growth rate $\beta$ are perturbed continuously as functions of $q$ according to $\beta(q)+\mathrm{j}\omega(q)=\beta_0(q)+\mathrm{j}\omega_0(q) + \lambda \mathcal{D}(q)$, where $\mathcal{D}$ is determined by nondegenerate perturbation theory. On the other hand, degenerate perturbation theory must be applied when $qN_h \mathrm{~mod~} \pi = 0$, which will mix the modes. The leading order perturbation in the degenerate case is $\beta(q)+\mathrm{j}\omega(q)=\beta_0(q)+\mathrm{j}\omega_0(q)+ \lambda\big(\mathcal{D}(q)\pm\sqrt{\mathcal{G}(q)}\big)$, where $\mathcal{D}$ and $\mathcal{G}$ are now determined by degenerate perturbation theory. Thus, heterogeneity will lift the degeneracy in the frequencies and growth rates, giving rise to a discontinuous jump of $2\sqrt{\mathcal{G}(q)}$ as a function of $q$, which will in general correspond to a band gap. The wavenumber $q$ does not reflect the symmetry of the heterogeneous system, since different values of $q$ can lead to identical eigenvalues for $T^m_{n}$. The wavenumber $k \equiv (qN_h \mathrm{~mod~} \pi)/N_h$, on the other hand, does account for this symmetry and is the parameterization that we employ throughout. Under this parameterization, the dispersion relation consists of $N_h$ wave branches that differ according to the mode dynamics within the $N_h$-unit cell and that separate from each other when the degeneracy is lifted. Band gaps will appear generically for frequencies between the branches (except potentially for the anomalous case of a nonmonotonic dispersion relation function $\omega_0(q)$).

When the system is parametrically driven with a driving frequency $\omega_d$ and amplitude $a_d$, the growth rate $\beta$ will increase rapidly with increasing $a_d$ for modes with frequencies $\omega=\omega_d/2$ that are resonant with the driving. For driving frequencies corresponding to twice the frequencies within a band gap, no resonant modes exist and larger driving amplitudes are required to induce instabilities (i.e., to turn $\beta$ positive for a nonresonant mode), resulting in HSHS.

 \section*{Faraday instability theory and numerics.}
\ We use Cartesian coordinates, with $(x,y)$ for the horizontal plane and $z$ for the vertical axis, and consider a fluid filling a container up to a height of $h_0$ when undisturbed. Given a surface deflection $h$ for each $(x,y)$, the surface is then located at $z=h_0+a_d \sin(\omega_d t)+h$ when driven by a vertical vibration. The surface deflection $h$ along with the fluid velocity $(u_x, u_y, u_z)$ and the pressure $P$ at each point $(x,y,z)$ completely specify the state of the system.

For simplicity, we assume the fluid is inviscid and incompressible. The undisturbed surface solution is given by $u_x=u_y=0$, $u_z=a_d\omega_d\cos(\omega_d t)$, $P=-gz$, and $h=0$. This is always a solution to the fluid equations of motion, and the question to be considered is the stability of this solution. Deviations from the undisturbed surface solution are described by velocity potential $\phi(x,y,z',t)$, where we have changed to an accelerated reference frame given by $z'=z+a_d\sin(\omega_d t)$. The velocity potential satisfies the Laplace equation with Dirichlet boundary conditions specified by a function $\zeta(x,y,t)$ on the free surface, which is located at $z'=h_0+h(x,y,t)$. For points $\mathbf{x}=(x,y,z')$ at the substrate bottom or at the side walls (which we denote as the set $\Omega$), the velocity potential satisfies Neumann boundary conditions. Thus, the equations determining the velocity potential are
\begin{align}
\nabla^2\phi = 0, ~
\left.
\phi \right|_{z'=h_0+h} = \zeta, \text{ and} \left. \hat{\mathbf{n}}\cdot \nabla \phi \right| _{\mathbf{x} \in \Omega} = 0, \label{laplace}
\end{align}
where $\hat{\mathbf{n}}$ is the unit normal vector pointing outward from the fluid and into the solid surfaces. The evolution of $\zeta$ follows from the Ber\-nou\-lli equation for inviscid flows
\begin{align}
\label{bernoulli}
\partial_t \zeta &=
\frac{\sigma}{\rho} \nabla \cdot \hat{\mathbf{n}}
- \big[ g-a_d\omega_d^2 \cos(\omega_dt) \big] h \nonumber \\
&\quad -\left.\Big[\frac{1}{2} \left|\nabla \phi\right|^2- \partial_{z'} \phi \partial_t h \Big]\right|_{z'=h_0+h,}
\end{align}
where $\rho$ is the fluid density and $\sigma$ is the surface tension. The evolution of the surface height $h$, on the other hand, is determined by the kinematic equation
\begin{align}
\label{kinematic}
\partial_t h = \left.\left[\partial_{z'} \phi- \nabla \phi \cdot \nabla h - \nu_1 h + \nu_2 \nabla^2 h \right] \right|_{z'=h_0+h},
\end{align}
where the term $- \nu_1 h + \nu_2 \nabla^2 h$ is included in order to mimic the neglected viscosity and regularize the numerics. The values of the damping parameters $\nu_1$ and $\nu_2$ are chosen to fit the experimental instability boundaries. The damping in experiments is dominated by the contact line pinning and cannot be easily predicted, so initial calibration experiments were run to estimate these parameters. The critical frequencies and accelerations for the three tongues in Fig.~\ref{fig4}a were measured, and the values $\nu_1=2.0$ \si{Hz} and $\nu_2=0.1$ \si{cm^2 / s} were found to give satisfactory fits. Since these damping parameters are expected to vary with the experimental setup and are likely not accurate outside of the given frequency range, these empirical estimates are considered adequate for our purposes. Equations \eqref{bernoulli}-\eqref{kinematic} are integrated numerically, where $\phi$ is computed from moving mesh finite element solutions of \eqref{laplace} each time step using the FEniCS package in Python\cite{2012_Logg}. In these simulations, we implemented two- and three-dimensional rectangular geometries for both periodic boundary conditions and pinned-contact-line boundary conditions. The substrate shapes in all cases can be specified as flat, periodic, or random.

As in the analysis of the pendulum array model, for the flat and sinusoidal substrates with periodic boundary conditions, the linearized equations of motion can be reduced to an eigenvalue problem, albeit a more complicated one in this case. This is accomplished by taking the Floquet-Fourier ansatz
\begin{align}
\phi = {\mathrm e}^{\mathrm{j}(kx+st)}\sum_i \sum_m \hat{\phi}_{im}{\mathrm e}^{\mathrm{j}ik_sx+\mathrm{j}m\omega_dt}, \\
\zeta={\mathrm e}^{\mathrm{j}(kx+st)}\sum_i \sum_m \hat{\zeta}_{im}{\mathrm e}^{\mathrm{j}ik_sx+\mathrm{j}m\omega_dt}, \\
h={\mathrm e}^{\mathrm{j}(kx+st)}\sum_i \sum_m \hat{h}_{im}{\mathrm e}^{\mathrm{j}ik_sx+\mathrm{j}m\omega_dt},
\end{align}
where we recall that $k_s$ is the substrate wavenumber and $a_s$ is the substrate amplitude. Projecting the equations onto the Floquet-Fourier modes, and employing the integral representation of the Bessel function $J_i(\mathrm{j}w) = \frac{1}{2\pi}\int_{-\pi}^{\pi}{\mathrm e}^{\mathrm{j}i\tau + w\sin\tau}{\mathrm d}\tau$, we obtain the linear form
\begin{equation}
\label{linear}
\sum_i \sum_m {C}^{im}_{jn}\hat{h}_{im} = a_d \sum_i \sum_m {D}^{im}_{jn}\hat{h}_{im}
\end{equation}
after eliminating $\hat{\phi}_{im}$ and $\hat{\zeta}_{im}$. The tensor elements in \eqref{linear} are given by
\begin{align}
{C}^{i m}_{j n} &= {\mathrm e}^{-\kappa_i h_0}\Big\{ -(s+m\omega_d)^2 +\mathrm{j}(s+m\omega_d)(\nu_1+\kappa_i^2\nu_2) \nonumber \\
&\quad -\kappa_i\Big(g+\frac{\sigma}{\rho}\kappa_i^2\Big)\Big\} \times \Big\{ J_{i-j}(\mathrm{j}\kappa_ia_s)- \frac{\mathrm{j}k_sa_s}{2} \nonumber \\
&\quad \times \big[ J_{i-j+1}(\mathrm{j}\kappa_ia_s) +J_{i-j-1}(\mathrm{j}\kappa_ia_s) \big] \Big\} \delta_m^n \nonumber \\
&+ \quad {\mathrm e}^{\kappa_i h_0}\Big\{ -(s+m\omega_d)^2 +\mathrm{j}(s+m\omega_d)(\nu_1+\kappa_i^2\nu_2) \nonumber \\
&\quad +\kappa_i\Big(g+\frac{\sigma}{\rho}\kappa_i^2\Big)\Big\} \times \Big\{ J_{i-j}(-\mathrm{j}\kappa_ia_s)+\frac{\mathrm{j}k_sa_s}{2} \nonumber \\
&\quad \times \big[ J_{i-j+1}(-\mathrm{j}\kappa_ia_s) +J_{i-j-1}(-\mathrm{j}\kappa_ia_s)\big]\Big\} \delta_m^n
\end{align}
and
\begin{align}
{D}^{i m}_{j n} &= \frac{\omega_d^2\kappa_i{\mathrm e}^{-\kappa_i h_0}}{2}\Big(\delta_m^{n+1}+\delta_m^{n-1}\Big)\Big\{ J_{i-j}(\mathrm{j}\kappa_ia_s)- \frac{\mathrm{j}k_sa_s}{2} \nonumber \\
&\quad \times \big[ J_{i-j+1}(\mathrm{j}\kappa_ia_s) +J_{i-j-1}(\mathrm{j}\kappa_ia_s)\big]\Big\} \nonumber \\
&\quad - \frac{\omega_d^2\kappa_i{\mathrm e}^{\kappa_i h_0}}{2}\Big(\delta_m^{n+1}+\delta_m^{n-1}\Big) \Big\{ J_{i-j}(-\mathrm{j}\kappa_ia_s)+ \frac{\mathrm{j}k_sa_s}{2} \nonumber \\
&\quad \times \big[ J_{i-j+1}(-\mathrm{j}\kappa_ia_s) +J_{i-j-1}(-\mathrm{j}\kappa_ia_s)\big]\Big\},
\end{align}
where $\kappa_i = k + ik_s$.
As in the pendulum array case, the eigenvalue problem in \eqref{faradayeigen} follows when we set the growth rate to zero and the driving amplitude to $a_d^{~*}$, and we map indices $\mu \in \mathbb{Z}$ and $\nu \in \mathbb{Z}$ to pairs of indices $(i(\mu),m(\mu))\in \mathbb{Z}^2$ and $(j(\nu),n(\nu))\in \mathbb{Z}^2$ using any convenient bijection from $\mathbb{Z}$ to $\mathbb{Z}^2$, taking $\tilde{C}_{\mu \nu} \equiv {C}^{i(\mu)\ m(\mu)}_{j(\nu)\ n(\nu)}$, $\tilde{D}_{\mu \nu} \equiv {D}^{i(\mu)\ m(\mu)}_{j(\nu)\ n(\nu)}$, and $\tilde{h}_{\mu} \equiv \hat{h}_{i(\mu)\ m(\mu)}$. For the periodic and flat substrates, the instability boundaries determined by the eigenvalue problem in \eqref{faradayeigen} agree completely with the boundaries determined directly by finite element simulations. For the random substrates, the instability boundaries are calculated directly from the finite element simulations. In all cases, our analysis is focused on the first Brillouin zone by setting $s=0$ and $s = \pm \mathrm{j} \omega_d/2$ to determine the harmonic and subharmonic instability boundaries, respectively.

For the previously studied case of a flat substrate, the dispersion relation can be derived by linearizing \eqref{laplace}-\eqref{kinematic} with a modal solution $\phi=\Phi(\mathbf{k},t) \exp(\mathrm{j}\mathbf{k}\cdot\mathbf{x})$, $h=H(\mathbf{k},t)\exp(\mathrm{j} \mathbf{k}\cdot\mathbf{x})$, and $\zeta=Z(\mathbf{k},t)\exp(\mathrm {j} \mathbf{k}\cdot\mathbf{x})$ of wavenumber $\mathbf{k}$. This results in the Mathieu equation,
\begin{equation}
\partial_t^2 H= -k \tanh (k h_0) \big[g+ \sigma k^2 -a_d \omega_d^2 \sin (\omega_d t) \big]H, \label{mathieu}
\end{equation}
after eliminating $\Phi$ and $Z$ while ignoring the phenomenological damping terms. The dispersion relation $\omega^2 = k\left(g+\sigma k^2\right)\tanh\left( k h_0\right)$ then follows from \eqref{mathieu} by setting $a_d=\omega_d=0$. The exponential dependence encoded by $\tanh(kh_0)$ in the homogeneous case suggests that the onset of instabilities can only be significantly affected by heterogeneous substrates if the minimum thickness $h_m$ for the undisturbed fluid surface satisfies $h_m \lesssim O(1/k)$, as confirmed in our modeling. For the non-flat substrates, the dispersion relation is determined numerically from the spectrum of $C^{i0}_{j0}$.\\

\section*{Faraday instability experiments.}
\ We used a model VG-100-8 Vibration Test Systems (VTS) shaker to drive the fluid. A Tektronix signal generator (model AFG1022) was attached to an EMB amplifier (model EB3500PRO) to drive the VTS shaker. Polylactic acid substrates were printed using a Monoprice Maker Ultimate 3D printer. The fluid domain of the printed substrates measured $4$ \si{cm} long by $1$ \si{cm} wide, and all containers held the same volume of fluid as the container with the flat substrate, which was $0.5$ \si{cm} deep. The substrates were affixed with epoxy to a $2.5$ \si{cm} thick acrylic plate, which was laser cut for mounting. Another acrylic plate held an Analog Devices ADXL337 accelerometer with a $\pm 3g$ acceleration range. The plates were stacked with a spacer leaving a gap for the accelerometer. They were then bolted together onto the VTS shaker, and the accelerometer was connected to a Sparkfun Redboard (model DEV-13975), which was programmed to fit a sinusoidal curve to the acceleration in order to extract the driving acceleration amplitude. The containers were filled to the brim to minimize the boundary effects\cite{1988_Douady_Fauve,1990_Douady}. The VTS shaker was leveled so that the gravitational field was normal to the fluid surface. A FLIR Blackfly S camera (model BFS-U3-32S4M-C) was synchronized with the driving frequency and used to track the growth of the surface deflection. For each driving frequency, starting from a small driving voltage (small acceleration in the VTS shaker) with a stable flat surface, the voltage of the signal generator was increased in $10$ \si{mV} increments, allowing approximately $5$ \si{s} to determine whether the flat surface would become unstable before increasing the voltage. When the instability appeared, the acceleration amplitude was recorded as the instability boundary for that frequency. The critical acceleration amplitude $\alpha_d^{~*}$ was converted to the critical driving amplitude $a_d^{~*}$ using $a_d^{~*}=\alpha_d^{~*}/\omega_d^2$. Frequencies were swept from $13$ \si{Hz} to $30$ \si{Hz} in increments of $0.5$ \si{Hz}, then back down to $13$ \si{Hz} again to collect a second data point. This process was repeated twice for each substrate, leading to $4$ data samples at each frequency.

 \end{methods}

\section*{Data availability.}
\ The data recorded during the Faraday instability experiments and shown in Fig.~\ref{fig4} are available from our GitHub repository:  \blue{\href{https://github.com/znicolaou/faraday}{https://github.com/znicolaou/faraday}}. The other data that support the findings of this study are computer generated (see code availability) and are available from the corresponding author upon reasonable request.

\section*{Code availability.}
\ The source code for determining instability boundaries, dispersion relations (including the ones not analytically expressed above), and gap soliton solutions is available from our GitHub repository:  \blue{\href{https://github.com/znicolaou/faraday}{https://github.com/znicolaou/faraday}}.


\section*{Acknowledgements\\}
This work was supported by the Complex Dynamics and Systems Program of the Army Research Office (Short-Term Innovative Research Grant No. W911NF-20-1-0173) and a Northwestern W Award. The authors thank P.\ B.\ Umbanhowar for providing the VTS mechanical shaker and V.\ Chandrasekhar and K.\ Ryan for facilitating the 3D printing of the substrates.
\section*{Author contributions\\}
Z.G.N., D.J.C., and A.E.M.\ designed the research. Z.G.N.\ and D.J.C.\ programmed simulations and performed numerics. Z.G.N.\ designed and conducted the experiments, E.B.W.\ guided the setup of the experimental apparatus, and M.M.D.\ provided lab space and equipment. Z.G.N.\ and A.E.M.\ analyzed the data and led the writing of the manuscript. The final presentation was approved by all authors.
\section*{Competing interests\\}
The authors declare no competing interests.
\pagebreak
\section*{Additional information}
\subsection{Supplementary information} The online version contains supplementary material at \href{https://doi.org/10.1038/s41467-021-24459-0}{\blue{https://doi.org/10.1038/s41467-021-24459-0}}.
\subsection{Correspondence} and requests for materials should be addressed to A.E.M.
\vfill

\end{document}


\renewcommand{\thefigure}{{\small \textbf{\arabic{figure}}}}
\renewcommand{\theequation}{{\normalsize \arabic{equation}}}
\renewcommand{\figurename}{\textbf{Supplementary Fig.}}
\makeatletter
\c@secnumdepth=1

\let\oldaddcontentsline\addcontentsline
\renewcommand{\addcontentsline}[3]{}
\title{
{\bf  {SUPPLEMENTARY INFORMATION}}\\[1em]{\it  Heterogeneity-stabilized homogeneous states in driven media}\vspace{-0.5em}}
\author{ Zachary G. Nicolaou}
\author{ Daniel J. Case}
\author{ Ernest B. van der Wee}
\author{ Michelle M. Driscoll}
\author{ Adilson E. Motter}
\maketitle
\let\addcontentsline\oldaddcontentsline
\onecolumngrid

\section*{Supplementary Discussion 1}
\subsection*{Band gaps in a pendulum array with heterogeneous masses}
Here, we consider an alternative pendulum model in which the masses of the pendula vary rather than the lengths. This case is {of easy experimental implementation and illustrates the generality of band gap opening} by periodic heterogeneities (see Methods). The equations of motion {for the driven array} are given by
\begin{equation}
M_i L \ddot{\theta}_i = -\eta L \dot{\theta}_i - M_i {[g-a_d\omega_d^2\cos(\omega_d t)]} \sin(\theta_i) + \kappa L[\sin(\theta_{i+1}-\theta_i) +\sin(\theta_{i-1}-\theta_i)] {,} \label{mass}
\end{equation}
{where $1\leq i \leq N$.}
 {In the absence of driving ($a_d=0$), linearization around $\theta_i=0\ \forall i$} leads to an {expression in the form of} (9){:
}\begin{equation}
{M_i \ddot{\theta}_i + \eta \dot{\theta}_i + \sum_j \left[(M_i g/L+2\kappa) \delta^i_{j} - \kappa (\delta^{i+1}_j+\delta^{i-1}_j) \right] \theta_j=0,
}\end{equation}
{where the heterogeneity is determined by the differing $M_i$}. Supplementary Fig.~1 shows how wave modes split into distinct branches for a specific periodic heterogeneity with a three-particle unit cell,
\begin{equation}
M_i =
 \begin{cases}
1-\lambda & i \mathrm{~mod~} 3= 0, \\
1 & i \mathrm{~mod~} 3= 1, \\
1+\lambda & i \mathrm{~mod~} 3= 2,
\end{cases}
 \label{mass2}
\end{equation}
with $L=1$, $g=1$, and $\kappa=1$. Similar to the {pendulum array} discussed in the main text, when this system is parametrically driven with driving frequencies around twice the frequencies {in a} band gap, there are no resonant wave modes that can be easily excited, and thus heterogeneity-stabilized homogeneous states emerge.
\clearpage
\begin{figure}[t]
\includegraphics[width=\columnwidth]{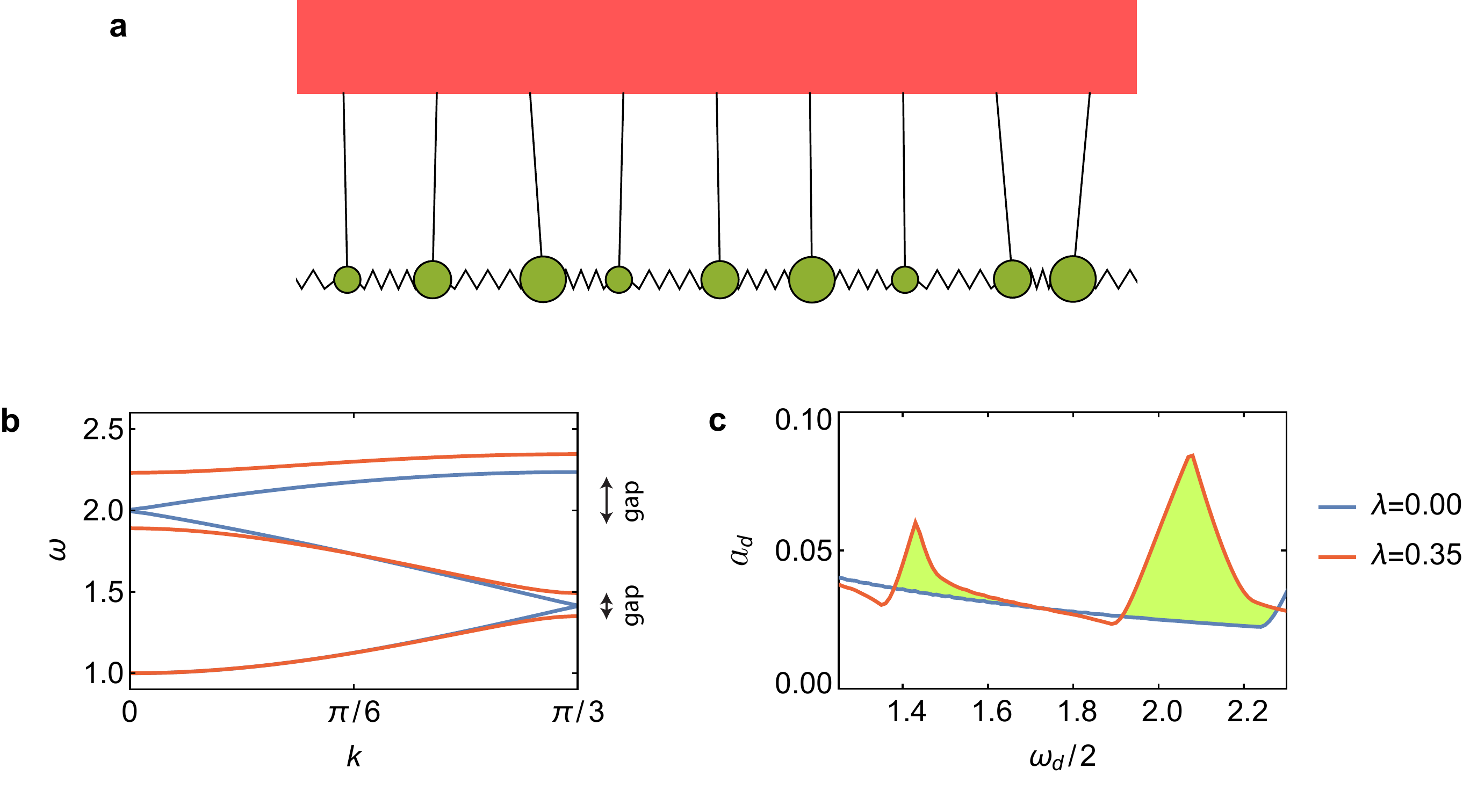}
\caption{\textbf{Band-gap opening and HSHS for the pendulum array with heterogeneous masses.} \textbf{a}, Schematic of the pendulum array with a three-particle unit cell described by Supplementary Eqs.~\eqref{mass}-\eqref{mass2}.  \textbf{b}, Frequency $\omega$ vs.\ wavenumber  {$k$} for {$\lambda=0$} ({blue} lines) {and $\lambda=0.35$} ({red} lines) {in the $N\to \infty$ limit for the system shown in \textbf{a}}. {Two gaps open between} the {three} branches {in} the {heterogeneous case}. {\textbf{c}, Instability boundaries for the systems shown in \textbf{b}, with green shading showing areas exhibiting HSHS associated with the respective band gaps in \textbf{b}.}
\label{figs1}}
\end{figure}

\section*{Supplementary Discussion 2}
\subsection*{Finite-size perturbations and gap solitons in driven pendulum {arrays}}
In the main text, we {established} that the homogeneous states {are stabilized in the band gaps} when heterogeneity is introduced, in the sense that infinitesimal perturbations around them decay. However, other stable states may {also} emerge when the system is driven.  
If multiple stable states coexist, sufficiently large perturbations {may induce transitions between states even when the initial state is} stable against infinitesimal perturbations. Here, we consider heterogeneity-stabilized homogeneous states in the presence of {finite-size} perturbations.  For concreteness, we focus on the driven pendulum array with periodic heterogeneity {considered} in the main text, which is defined by alternating pendulum length {$L_i=\overline{L} + (-1)^i\Delta$}.
 \begin{figure}[b]
 \includegraphics[width=\columnwidth]{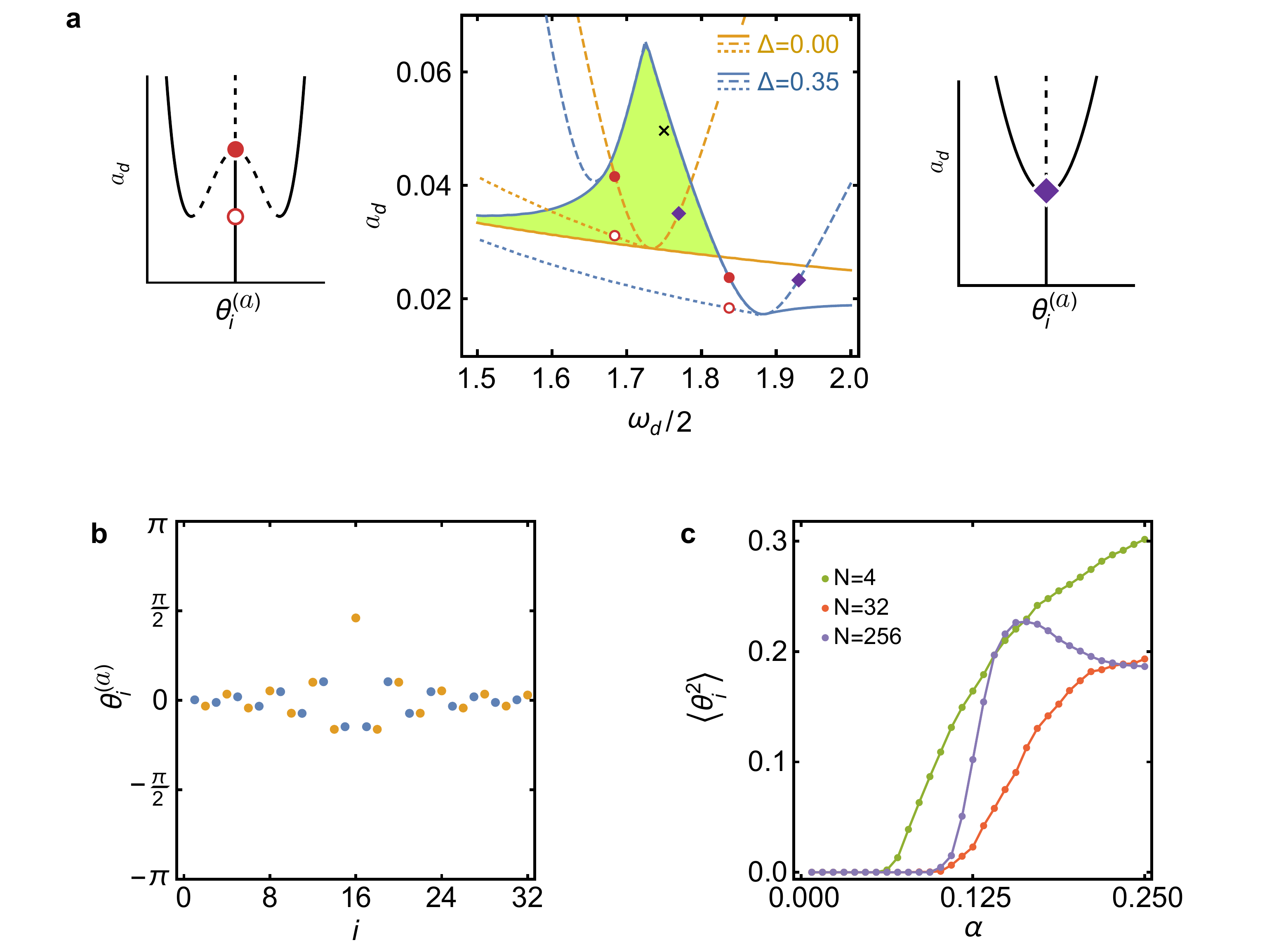}
 \caption{\textbf{Multistability and {finite-size} perturbations in pendulum array model.} \textbf{a}, Primary instability boundary  {for the homogeneous array} (solid {orange line}) and heterogeneous array with $\Delta=0.35$ (solid blue line) in the $N\to\infty$ limit. The boundaries for the $k=\pi/2$ instability mode (dashed {lines}) and the corresponding {secondary instability boundary} (dotted lines) are also shown, {along} with the region of HSHS (green area). Outsets: {driving amplitude vs.\ swinging amplitudes} for {negative detuning parameters}  (subcritical bifurcation, {left} panel) and {positive detuning parameters} (supercritical bifurcation, {right} panel).
 Solid lines show stable states, dashed lines show unstable states, and symbols (also marked in the central panel) indicate the instability boundary (filled circles and diamonds) and the secondary instability boundary (open circles) for a particular mode. \textbf{b}, {Swinging amplitude} vs.\ pendulum index for a localized gap soliton state in a heterogeneous array determined by the parameters marked by $\times$ in \textbf{a}, with short pendula shown as orange dots and long pendula shown as blue dots.  \textbf{c}, Average {(over time and perturbation realization) of the} squared phases of final stable state $\langle \theta_i^2 \rangle$ vs.\  perturbation {size} $\alpha$ for heterogeneous arrays of various size with the other parameters corresponding to $\times$ in \textbf{a}. \label{figs2}}
\end{figure}

{We} find that alternative stable states do, in fact, emerge in the periodic pendulum array in {parameter regions that} exhibit heterogeneity-stabilized homogeneous states, as shown in Supplementary Fig.~2. This follows because the instability {of any particular wave mode} becomes subcritical when the driving frequency is smaller than the resonant frequency of that mode {(left panel in Supplementary Fig.~2a), which should be contrasted with the supercritical form of the instabilities for larger driving frequencies (right panel in Supplementary Fig.~2a)}. {The difference between the driving frequency and the resonant frequency of a particular mode is called the detuning parameter for that mode, so modes exhibit subcritical instabilities} for negative detuning parameters. 
 {In the subcritical cases, an unstable swinging state vanishes at a \textit{secondary} instability boundary of a new stable periodic swinging state that emerges for increasing driving amplitude.} Since the detuning parameter for the mode is negative in this case, 
 there exists a {different,} lower frequency mode that is resonant with the driving frequency. 
{Thus,} the {secondary instability boundary} for the swinging states may lie above or below the {(primary)} instability boundary for the uniform state, which consists of the envelope of the instability boundaries over all modes.

{The central panel in Supplementary Fig.}~2a shows the {primary} instability boundaries for the pendulum array {in the} $N\to\infty$ limit with $\Delta=0$ and $\Delta=0.35$ (solid lines). The dashed lines show the instability boundary for the {$k=\pi/2$} mode that is split by the bad gap, the dotted lines show the {secondary instability boundaries for this mode} in the subcritical {cases}.  
For the homogeneous array ({orange lines}){, the secondary instability boundary for this mode} lies above the {primary} instability boundary, so the system is not susceptible to  {finite-size} instabilities below the {primary} instability boundary. For the heterogeneous array {in this figure} ({blue lines}), on the other hand, the {secondary instability boundary for this mode} lies below the {primary} instability boundary, and {finite-size} perturbations can therefore excite the array to {the} periodic swinging state below the {primary} instability boundary.
In the outsets, the $\theta_i^{(a)}$ describe the swinging state amplitudes, defined by the value of the phases at the time points where $\sum_i \theta_i^2$ is maximized.  

{Interestingly, we find that} the periodic swinging states are not the only alternative stable {states} for the heterogeneous array. Supplementary Fig.~2b shows {the swinging state amplitudes for} a particularly interesting localized stable {state that we observe for random initial conditions} in an array of $32$ pendula for the parameter values marked by the $\times$ ($\omega_d=3.5$ and $a_d=0.05$) and {heterogeneity in Supplementary Fig.~2a}. 
An animation of this localized state is also available {as part of Supplementary Movie 1. {Since these localized states can coexist in a variety of spatial configurations, the heterogeneous pendulum array exhibits a high degree of multistability}. Similar gap soliton solutions have been observed around band gaps in other media, emerging through a {snaking bifurcation in pattern-formation models} \cite{2018_PonedeL_Knobloch}. While beyond the scope of this work, we expect that such gap solitons will also exist in Faraday instability systems with periodic substrates within regions of HSHS{. We argue that this is expected because the bifurcations of instability modes} have been shown to become subcritical for negative detuning parameters in Faraday wave experiments {with homogeneous substrates} \cite{1993_Cross_Hohenberg}, {in direct analogy with the orange lines in Supplementary Fig.~2a}.

 {Finite-size} perturbations can induce a transition between the homogeneous state and {the stable swinging states below the instability boundary}. To quantify the stability against {finite-size} perturbations, we {simulate} the system with a random initial perturbation given by an initial $\theta_i$ uniformly distributed in $[-\alpha\pi, \alpha\pi]$ and an initial $\dot{\theta_i}=0$. Here, $\alpha$ quantifies the {size} of the perturbation. {The system is evolved} until it approaches a stable state. Supplementary Fig.~2c shows the  {time-averaged} value of the phases {(after the decay of the initial transient) averaged over} $1024$ random {perturbation realizations} for arrays of various sizes for the parameter values shown by the $\times$ in Supplementary Fig.~2a. The stability transition sharpens as the number of pendula increases and, for large $N$, the homogeneous state is stable against almost all {simulated} finite-amplitude perturbations for $\alpha < 0.1$. Thus, {finite-size}  perturbations can destabilize heterogeneity-stabilized homogeneous states and lead to nontrivial dynamical states, but doing so requires {large} perturbations.
 